\documentclass[runningheads]{llncs}
\usepackage{graphicx}
\usepackage[table,xcdraw]{xcolor}
\usepackage{cite}
\usepackage{amsmath,amssymb,amsfonts}
\usepackage{algorithmic}
\usepackage{graphicx}
\usepackage{textcomp}
\usepackage{import}
\usepackage{xcolor}
\usepackage{hyperref}
\usepackage{amsmath}
\usepackage{algorithm}  
\usepackage{algorithmic}

\usepackage{booktabs}

\usepackage{import}
\usepackage{cite}

\begin{document}

\title{Property-based Entity Type Graph Matching}

\author{Fausto Giunchiglia\  \orcidID{0000-0002-5903-6150} 
\and Daqian Shi\  \orcidID{0000-0003-2183-1957} } 
%
%

\authorrunning{F. Giunchiglia and D. Shi}

\institute{Department of Information Engineering and Computer Science (DISI),\\ University of Trento, Italy\\
\email{\{fausto.giunchiglia, daqian.shi\}@unitn.it} 
}

\maketitle

\begin{abstract}
We are interested in dealing with the heterogeneity of Knowledge bases (KBs), e.g., ontologies and schemas, modeled as sets of entity types (etypes), e.g., person, where each etype is associated with a set of properties, e.g., age or height, via an inheritance hierarchy. A huge literature exists on this topic. A common approach is to model KBs as graphs decorated with labels and reduce the problem of KB matching to that of matching these two elements, \textit{viz.}, labels and structure of the graph. 
However, labels of etypes are often misplaced, e.g., they are more general or specific than the correct etype, as defined by its properties. Structure-based matching may also lead to wrong conclusions as the properties assigned to an etype in a inheritance hierarchy do not depend on the order by which they are assigned and, therefore, on the specific structure of the graph. 
In this paper, we propose a novel etype graph matching approach, dealing with the two problems highlighted above, based on two key ideas. The first is to implement matching as a classification task where etypes are characterized by the associated properties. The second is we propose two \textit{property-based} etype similarity metrics, which model the roles that properties have in the definition of an etype. The experimental results show the effectiveness of the algorithm, in particular for those etype graphs with a high number of properties.

\end{abstract}

\section{Introduction}
We are interested in dealing with the heterogeneity of Knowledge bases (KBs), e.g., ontologies and schemas, modeled as sets of entity types (etypes), e.g., person, where each etype is associated with a set of properties, e.g., age or height, via an inheritance hierarchy. A huge literature exists on this topic, e.g., \cite{2,3,39}. Most etype graph matching approaches exploit label-based methods \cite{18,19}, such as character similarity metrics and synonym analysis, and structure-based methods \cite{46}, implementing various forms of graph matching. 
However, labels of etypes may suggest a wrong etype \cite{8,9}. For example, an eagle can be labelled as \textit{Bird} in a general-purpose ontology and   \textit{Eagle} in a domain-specific ontology. Structure-based matching may also lead to wrong conclusions as the properties assigned to an etype in an inheritance hierarchy are cumulative and depend only on the nodes in the path from the root and, therefore, do not depend on the order by which they are assigned. For example, the super-class of etype \textit{Eagle} can be \textit{Animal} in one etype graph and \textit{Bird} in another etype graph. 

As a solution to the above problems, the main intuition underlying the work described in this paper is to match etypes on the basis of the properties which are used to define them. It is, in fact, the properties that are used to intensionally define an etype which define it independently of its specific name and also independently of its hierarchy \cite{10}. Furthermore, it is fact that in most relevant ontologies, etypes are associated with sufficient properties, like DBpedia \cite{11} and OpenCyc \cite{45}. And the reason for this is quite obvious, being the purpose of any data or knowledge integration task exactly that of extending the number of properties associated to an etype. 

In this paper, we implement the above intuition based on main contributions:
\begin{itemize}
    \item We introduce two \textit{property-based} etype similarity metrics, namely the \textit{horizontal similarity} $ES_h$ and the \textit{vertical similarity} $ES_v$ which characterise the role that properties have in the definition of given etypes. These similarity metrics capture the main idea that for any two etypes, the properties which distinguish one etype from the other should not occur in the other etype. Since different properties contribute differently for matching etypes, we introduce $ES_h$ which focuses on measuring the properties with different shareability, and $ES_v$ measures properties based on their specificity. 
    \item We implement the etype graph matching as a classification task where the matching of etypes is based on their associated properties. In this paper, we propose and evaluate a machine learning (ML)-based etype graph matching approach.
    
\end{itemize}




The  paper is organized as follows. Section 2 introduces our own specific formalization for etype graphs and relevant terminology. Section 3 presents two property-based etype similarity metrics. Section 4 introduces the overall etype graph matching algorithm. The evaluation details and results in Section 5, where the experiments are based on a selected test cases from the Ontology Alignment Evaluation Initiative (OAEI) \cite{13}. Finally, we present the related work in Section 6 and the conclusions in Section 7.

\section{Etype Graphs as FCA contexts}
\label{sec3}
We formalize etype graphs as formal concept analysis \cite{10} (FCA) contexts. Specifically, we define an etype graph $ETG$ as $ETG = \langle E,P,T \rangle $, with $E = \{e_1…e_n\}$ being the set of etypes from the etype graph, $P = \{p_1…p_n\}$ being the set of properties, $T = \{e \in E |\langle e,T(e) \rangle \}$ being the set of correspondences between etypes and properties, where function $T(e)$ returns properties of $e$. We consider the property $p$ is used to describe an etype $e$ when the property belongs to set $T(e)$. Two observations:
\begin{enumerate}
    \item $E$ is a set of etypes but not a set of entities. Similar to what happens in general FCA, which assumes that an entity is described by a set of property values, an etype is considered to be described by a set of properties $T(e)$. Since in our method we focus on the correlations between etypes and properties, we organize an etype graph as etype-property correlation map as an FCA context without containing additional information.
    \item Etype characterization exploits not only the properties associated with it but also the others, namely those which are not used in its definition. Thus, we introduce the non-associated properties into our FCA context and distinguish two more different cases for better presenting the FCA context. 
\end{enumerate}

\begin{figure}[htbp]
	\centering
	\includegraphics[scale=0.5]{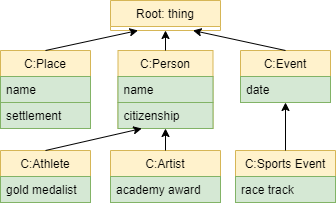}
	\caption{An example the hierarchy of etype graph \label{fig:2}}
\end{figure}
\noindent
As an example, Figure 1 presents the hierarchy of an etype graph, extracted from DBpedia \cite{11}. In each box, etypes are presented in yellow and their properties in green. We formalize the etype graph in Figure 1, into an FCA context as from below.

\begin{figure}[htbp]
	\centering
	\includegraphics[scale=0.6]{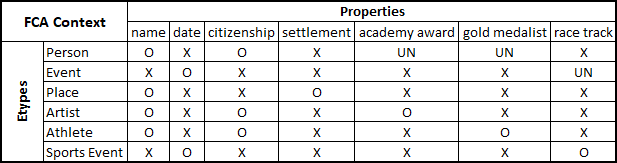}
	\caption{An example of formalizing etype graph into FCA contexts \label{fig:1}}
\end{figure}
\vspace{-0.5cm}
\noindent
In Figure 2 we adopt the following conventions. The value box with a circle represents the fact the property is associated with the etype, e.g., \textit{citizenship} is associated with \textit{Person}. The value box with a cross means the property is not associated with the etype, e.g., \textit{date} is not used to describe etype \textit{Person}. The value “UN” represents the fact that the property is not associated with the etype but associated with at least one of its subclasses, namely undefined. The intuition is that the property might or might not be used to describe the current etype, e.g., \textit{academy award} is used to describe \textit{Artist} and it might be used to describe \textit{Person} since \textit{Artist} is a subclass of \textit{Person}. We encode these three correlations as the parameter $w_p$. Since the correlation of “associated with” is positive for a property describing an etype, the correlation of “not associated with” is negative and the correlation of “undefined” is neutral, we take $w_p$ to be defined as $w_p \in \{1,0,-1\}$.

\begin{equation}
\small
w_p = \left\{
\begin{tabular}{ll}
1,  & if $p \in prop(E)$\\
0,  & if $p \notin prop(E) \& p \in prop(E.subclass)$ \\
-1, & if $p \notin prop(E) \& p \notin prop(E.subclass)$ 
\end{tabular}\right.
\end{equation}
In the above equation, we take $p$ as the target property and $prop(E)$ as the properties associated with $E$. Thus, the circles, UNs and crosses in Figure 2 are set to 1, 0 and -1, respectively.

\section{Property-based similarity}
\label{sec4}
The similarity metrics are inspired to the work in \cite{9,48} in considering properties as one of the most important features to describe an etype and to the formalization of the “get-specific” heuristic provided in \cite{12}. These provide us the intuition that a more specific property provides more information to identify an etype. Let us introduce our two etype similarity metrics in detail.

\subsection{Horizontal Similarity}

When measuring the specificity of a property, a possible idea is to horizontally compare the number of etypes that are described by a specific property, namely the shareability of the property \cite{9}. If a property is used for describing diverse etypes, it means that the property is not highly characterizing. Thus, for instance, in figure \ref{fig:1}, the property \textit{name} is used to describe \textit{Person}, \textit{Place}, \textit{Athlete}. Dually, if a property is used for describing a few etypes or the property is associated with only one etype, this means this property can be regarded as highly characterizing, e.g., in Figure \ref{fig:1}, property \textit{settlement} is specific for etype \textit{Place}. 
Based on this intuition, we consider the specificity of a property is related to its shareability. Therefore, we propose $SP$ as the metric for measuring property specificity. More precisely, $SP$ aims to minimize the number of etypes that are associated with the target property in a specific etype graph. We model the metric $SP$ as: 

\begin{equation}
\small
SP_{ETG}(p)= w_p * {e^{\lambda(1-n(p))}} \in [-1, 1]
\end{equation}
where $p$ is the input property and $n(p)$ is the number of etypes that are described by the input property in a specific entity graph $ETG$, thus $n(p)\ge0$; $e$ refers to the natural mathematical constant \cite{15}; $\lambda$ is a constraint factor whose aim  is to produce a gentle curve. 
%
Assume that $A$ and $B$ are two etype graphs. Then we model $ES_h$ as follows: 

\begin{equation}
\small
ES_{h}(E_a,E_b) =\frac{1}{2}\sum_{i=1}^{k} \left ( \frac{SP_{A}(p_i)}{|prop(E_a)|} + \frac{SP_{B}(p_i)}{|prop(E_b)|} \right) \in [0, 1]
\label{equ:3}
\end{equation}

\noindent where we take $E_a$, $E_b$ as the candidate etypes from $A$ and $B$ respectively. Thus $E_a \in A$ and $E_b \in B$; $prop(E)$ refers to the properties associated with the specific etype and $|prop(E)|$ refers to the number of $prop(E)$. $k$ is the number of matched properties which are associated with both etype $E_a$ and $E_b$. $SP_{A}(p_i)$ and $SP_{B}(p_i)$ refer to the specificity of the aligned property $p_i$ in $A$ and $B$, respectively. Notice that we have $ES_{h}(E_a,E_b) = ES_{h}(E_b,E_a)$. Notice also that we apply z-score normalization \cite{38} to $ES_h$ at the end of calculation, and that the range of $ES_h$ is between $0$ to $1$. 

\subsection{Vertical Similarity}

Etype graphs are organized as classification hierarchies such that upper-layer etypes represent more abstract or more general concepts, whereas lower-layer etypes represent more concrete or more specific concepts \cite{12,16}. Correspondingly, properties of upper-layer etypes are more general since they are used to describe general concepts, vice versa, properties of lower-layer etypes are more specific since they are used to describe specific concepts. We assume that specific properties will contribute more to the identification of an etype. For instance, in Figure \ref{fig:1}, as a lower-layer etype, \textit{Artist} can be identified by the property \textit{academy award} but not by the property \textit{name}. 
Based on this intuition, we propose L(p) as a metric for measuring  property specificity. We model L(p) as follows:

\begin{equation}
\small
L_{ETG}(p)= w_p * \theta * \min_{E \in etype(p)} layer(E) \in [-1, 1]
\end{equation}
where: $\theta$ is a constraint factor which normalized the range of the function; $etype(p)$ outputs all the etypes that are described by the property $p$; and $layer(E)$ refers to the layer of the inheritance hierarchy where an etype $E$ is defined. We define the vertical etype similarity metric $ES_v$ as from below.

\begin{equation}
\small
ES_{v}(E_a,E_b) =\frac{1}{2}\sum_{i=1}^{k} \left ( \frac{L_{A}(p_i)}{|prop(E_a)|} + \frac{L_{B}(p_i)}{|prop(E_b)|} \right) \in [0, 1]
\label{equ:5}
\end{equation}
\noindent
 Similar to the definition of $ES_h$, we have candidate etypes  $E_a \in A$ and $E_b \in B$ and the properties $prop(E)$ associated with the etype $E$. The key difference is that $ES_v$ exploits the property specificity based on the layer information $L(p)$. $L_{A}(p_i)$ and $L_{B}(p_i)$ refer to the highest layer of the aligned property $p_i$ in $A$ and $B$, respectively. Notice that $ES_v$ is symmetric as well. $ES_v$ is also normalized by z-score normalization, in the same way as  $ES_h$. Finally the range of $ES_v$ is between $0$ to $1$.

\section{Etype Graph Matching}
\label{sec5}

Figure \ref{fig:2} presents the Processing chart of our etype graph matching approach. It mainly consists of two matchers, the property matcher and the etype matcher. After parsing the input etype graph pair, properties are first sent into the NLP-based property matcher, where string-based and language-based similarity metrics are exploited to match two property labels \cite{51,53}. Then we generate the FCA contexts according to the etypes and correlated property pairs. In this phase, we will also generate our property-based etype similarity metrics $ES_h$ and $ES_v$ and then send them all to the etype matcher. We develop a ML-based matcher which considers etype matching as a binary classification task. Thus, our etype matcher will predict two incoming etypes as match or unmatch and output the matched etypes as the final results. 

\begin{figure}[htbp]
	\centering
	\includegraphics[scale=0.5]{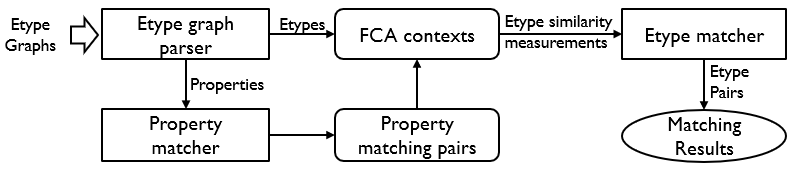}
	\caption{Processing chart of our etype graph matching approach\label{fig:2}}
\end{figure}

Algorithm 1 below presents the step-by-step process for calculating property-based etype similarity metrics $ES_h$ and $ES_v$. After formalizing etype graphs into FCA contexts, we assume that the two candidate FCA contexts $f_a$ and $f_b$ are generated. $PM$ refers to the property pairs which are aligned by the property matcher, $EM$ refers to the candidate etype pairs which are waiting for matching. For every etype pair in $EM$, we check their correlated properties and update the specificity values to $ES_h$ or $ES_v$ if the property pair is aligned. After traversing all the candidate etype pairs, we obtain completed etype similarities which will be used for training the ML model, or predicting if two etypes are matching. Table \ref{tab3} provides some representative examples to show the etype similarity $ES_v$ and $ES_h$ between etypes from \textit{cmt-confof} and \textit{cmt-conference} in conference track. 

\begin{algorithm}[htb] 
\caption{Etype similarity generation. $ES_h,ES_v=etypesim(f_a,f_b)$} 
\label{alg:1} 
\begin{algorithmic}[1] 
\REQUIRE ~~\\ 
Candidate FCA contexts $f_a$ and $f_b$;\\
\ENSURE ~~\\ 
Property-based etype similarity $ES_h,ES_v$;\\

\STATE 	$PM = (p_a, p_b) = PropertyMatcher(f_a,f_b)$; \{align $p_a$ and $p_b$ as property pairs by property matcher, where $p_a \in f_a$ and $p_b \in f_b$.\}
\STATE $EM = (E_a,E_b) = EtypeSelector(f_a,f_b)$; \{select etypes $E_a, E_b$ from $f_a, f_b$ and assemble them as candidate etype pairs $EM$.\}
\FOR {all $EM_i \in EM$}
\FOR {all $p_a \in f_a, p_b \in f_b$}
\IF{$(p_a,p_b) \in PM$}
\STATE $ES_h(EM_i).add( SP(p_a), SP(p_b))$; \{add the horizonal specificity to etype similarity $ES_h(EM_i)$, refers to equation \ref{equ:3}\}
\STATE $ES_v(EM_i).add( L(p_a), L(p_b))$; \{add the vertical specificity to etype similarity $ES_v(EM_i)$, refers to equation \ref{equ:5}\}
\ENDIF
\ENDFOR
\ENDFOR
\RETURN $ES_h, ES_v$
\end{algorithmic}
\end{algorithm}

\begin{table}[htb]
\centering
\vspace{0.2cm}
\caption{\textit{Examples of values of etype simimlarity $ES_v$ and $ES_h$}}
  \label{tab3}
 \vspace{-0.2cm}
\begin{tabular}{|l|l|l|l|}
\hline
\rowcolor[HTML]{FE996B} 
etype-cmt   & etype-confof         & $ES_v$ & $ES_h$ \\ \hline
Paper       & Contribution         & 1      & 0.853  \\ \hline
Author      & Author               & 0.756  & 0.740  \\ \hline
SubjectArea & Topic                & 0.198  & 0.961  \\ \hline
Meta-Review & Poster               & 0      & 0.312  \\ \hline
\rowcolor[HTML]{FE996B} 
etype-cmt   & etype-conference     & $ES_v$ & $ES_h$ \\ \hline
Chairman    & Chair                & 1      & 0.559  \\ \hline
Person      & Person               & 1      & 0.970  \\ \hline
Person      & Conference\_document & 0.02   & 0.06   \\ \hline
Chairman    & Publisher            & 0      & 0.07   \\ \hline
\end{tabular}
\end{table}


\section{Evaluation}
\label{sec6}
We first describe the evaluation set-up and then provide the results from the experiments.

\subsection{Evaluation Set-up}

The main decision for the evaluation was to take OAEI as the main reference for the selection of the matching problems. As of today, this in fact the major source of ontology matching problems. 

Our approach focuses on ontologies that contain etypes associated with a fair number of properties. As a result, we have selected the following cases: the bibliographic ontology dataset \cite{26} and conference track \cite{25} ($ra1$ version). From the bibliographic ontology dataset, we select series \#101 and series \#301-304, which present real-life ontologies for bibliographic references from the web. We set these bibliographic ontologies as the training set for training our ML-based etype matcher. The conference track contains 16 ontologies, dealing with conference organizations, and 21 reference alignments. We set the 21 reference alignments from the conference track as the testing set to validate our etype matcher. We select the training and testing set from different cases since we aim to prove the adaptation of our approach, which also prevents our approach from overfitting. Notice that there is an unbalanced positive and negative sample issue when we match two candidate ontologies, which means negative samples will be produced much more than positive samples. To address this issue, we propose a model training strategy that decreases the negative samples and duplicates a part of positive samples to achieve a balanced training set and to alleviate overfitting.

In this paper, our matching approach applies a general binary classification strategy, which is independent of the specific ML model. Thus, the data label is 1 or 0, which means two etypes are matching or unmatching respectively. The data consists of three kinds of attributes, which are string-based similarity metrics (N-garm \cite{29}, Longest common subsequence \cite{29}, Levenshtein distance \cite{30}), language-based similarity metrics (Wu and Palmer similarity \cite{31}, Word2vec \cite{32}) and property-based similarity metrics ($ES_h$ and $ES_v$). These etype similarities aim to measure different aspects of the relatedness between two etypes. Here we select some of the most common string-based and language-based similarity metrics as additional metrics working with our property-based similarity metrics for achieving better etype matching results.

\subsection{Experimental Results}

 For better evaluating the validity of our approach, we apply 4 different ML models, which are: random forest \cite{33}, stochastic gradient descent (SGD) classifier \cite{34}, decision tree \cite{35} and logistic regression \cite{36}. We have compared our work with state-of-the-art matching methods, as they came out of previous OAEI evaluation campaigns. 
The involved state of the art systems are: FCAMap \cite{6}, AML \cite{22}, LogMap and LogMapLt \cite{7}. We calculate precision, recall, $F_1$-measure, and also $F_{0.5}$-measure and $F_2$-measure \cite{37}.

\begin{table}[htbp]
\centering

\vspace{0.2cm}
\caption{\textit{Comparing our method with SOTA methods}}
  \label{tab1}
 \vspace{-0.2cm}
 
\begin{tabular}{cccccc}
\toprule
\textbf{ConferenceTrack}& \multicolumn{1}{l}{\textbf{ Prec. }} & \multicolumn{1}{l}{\textbf{ Rec. }} & \multicolumn{1}{l}{ \textbf{$F_{0.5}$}\textbf{-m.} } & \multicolumn{1}{l}{ \textbf{$F_1$}\textbf{-m.} } & \multicolumn{1}{l}{ \textbf{$F_2$}\textbf{-m.} } \\ \midrule
\textbf{FCAMap}                  & 0.680                                  & 0.625                               & 0.668                                  & 0.651                                 & 0.635                                 \\
\textbf{AML}                     & \textbf{0.832}                         & 0.630                               & \textbf{0.782}                         & \textbf{0.717}                        & 0.662                                 \\
\textbf{LogMap}                  & 0.798                                  & 0.592                               & 0.746                                  & 0.680                                 & 0.624                                 \\
\textbf{LogMapLt}                & 0.716                                  & 0.554                               & 0.676                                  & 0.625                                 & 0.580                                 \\
\textbf{Ours-RandomForest}       & 0.529                                  & \textbf{0.884}                      & 0.575                                  & 0.662                                 & \textbf{0.779}                        \\
\textbf{Ours-SGDClassifier}      & 0.779                                  & 0.632                               & 0.744                                  & 0.698                                 & 0.656                                 \\
\textbf{Ours-DecisionTree}       & 0.671                                  & 0.703                               & 0.677                                  & 0.687                                 & 0.696                                 \\
\textbf{Ours-LogisticRegression} & 0.556                                  & 0.808                               & 0.593                                  & 0.659                                 & 0.741         \\ \bottomrule                    
\end{tabular}
\end{table}
\noindent
Table \ref{tab1} shows the results of our approach with the different models mentioned above, compared  with the results of state-of-the-art methods. Firstly, we can find our approach with different models produce slightly different results, the SGD classifier performs the best in general, which leads the precision, $F_{0.5}$-measure and $F_1$-measure. And, random forest advances in recall and $F_2$-measure. Decision tree and logistic regression classifiers are marginally powerless than the other two in conference track ontologies. Secondly, extend to the overall comparison, we can find that AML has the best overall results. Leading the precision, $F_{0.5}$-measure and F1-measure. Our approach with random forest leads the results on recall and $F_2$-measure. Considering that the average results of our approach with different models are performing close to the state-of-the-art on $F_1$-measure, we can say that our approach leads to similar results as state-of-the-art competitors, while advances in different aspects\footnote{All approaches do not have significant differences in running times since the conference track contains no large ontology.}. 

The comparison to state-of-the-art methods shows the validity of our etype matcher. Moreover, we design a second experiment which is an ablation test to evaluate if our designed property-based etype similarity metrics are effective. In this experiment, we test on the backbone model (B) which was trained only by string-based and language-based similarity metrics. We also test on the model with $ES_h$, $ES_v$, and both $ES_h$ and $ES_v$ (ours), respectively. Note that the backbone model refers to \textit{Ours-SGDClassifier} in table \ref{tab1}.

Table \ref{tab2} shows the results of the ablation test, it is easy to find by using our designed metrics, the results significantly improved comparing with the results of the backbone model. Moreover, although B+$ES_h$ achieve the best recall measure, B+$ES_v$+$ES_h$ leads in precision and $F_1$-measure, which means the best overall performance. This observation shows both our designed metrics are effective on the etype matching task. At the same time, the etype matcher achieves the best performance by simultaneously using $ES_v$ and $ES_h$. 

\begin{table}[htbp]
\centering

\vspace{0.2cm}
\caption{\textit{Ablation test}}
  \label{tab2}
 \vspace{-0.2cm}
 
\begin{tabular}{@{}cccc@{}}
\toprule
\textbf{Models}        & \textbf{Prec.} & \textbf{Rec.}  & \multicolumn{1}{l}{\textbf{$F_1$}\textbf{-m.}} \\ \midrule
B                      & 0.621          & 0.605          & 0.613                             \\
B+$ES_v$               & 0.650          & 0.700          & 0.674                             \\
B+$ES_h$               & 0.634          & \textbf{0.729} & 0.678                             \\
B+$ES_v$+$ES_h$ & \textbf{0.779} & 0.632          & \textbf{0.698}                    \\ \bottomrule
\end{tabular}
\end{table}

\section{Related work}
\label{sec7}
Based on the idea originally introduced in \cite{9} and different from all the previous work, our approach is based on the idea of exploiting properties as the main means for matching etypes. We provide below a short summary of the four main techniques that we exploit in the implementation of property-based etype similarity, namely, label matching, graph matching, and the use of ML and FCA.

In the early stages of ontology matching, etype matching methods mostly focused on string-based methods. The work in \cite{19} reviews a wide range of string similarity metrics and propose an ontology alignment method by selecting different powerful similarity metrics. Later, ensemble metrics strategies were introduced in some studies \cite{24}, which apply multiple matchers based on different string-based metrics. The principle of these works is that the combined matchers are more powerful than an individual matcher.

The structure of an etype graph has also been considered as important information for identifying etypes, like \cite{46,52}. The LogMap system \cite{7} uses a two-step matching strategy, that is, matches two etypes $E_a$ and $E_b$ by a lexical matcher, and then considers the etypes that are semantically close to $E_a$ are more likely to be semantically close to $E_b$. AML \cite{22} introduces an ontology matching system that consists of a string-based matcher and a structure-based matcher, building internal correspondences by exploiting \textit{is-a} and \textit{part-of} relationships.

Some work on matching etypes is based on the use of ML. This work models the etype matching task as a binary classification task, trying to encode the information like string similarities and structure information as attributes. For instance, the work in \cite{21} achieves promising results by encoding the lexical similarity of the superclass and subclass as structural similarity. 

Finally, FCA lattices have been applied in etype matching methods in the work described in \cite{6,44}. To refine health records searching outputs, the work in \cite{42} introduced a matching method based on FCA which assists the end-user in defining their queries. In turn, in \cite{44} a bottom-up ontology merging approach was proposed where FCA lattices were used to keep track of the ontology hierarchy.

\section{Conclusions}
\label{sec8}
In this paper, we have introduced a novel etype graph matching approach via property-based similarity measurement. Firstly, we discussed a novel formalization method for etype graphs, which encodes etypes and properties into FCA contexts. Then we proposed two novel metrics for measuring the contextual similarity between two etypes, namely horizontal similarity and vertical similarity. Based on our proposed metrics, we have developed a ML-based framework for etype graph matching. The experimental results show the validity of our approach.

\section*{Acknowledgements}

The research conducted by Fausto Giunchiglia 
has received funding from the \emph{InteropEHRate} project, co-funded by the European Union (EU) Horizon 2020 programme under grant number 826106, and the research conducted by Daqian Shi has received funding from the program of China Scholarships Council (No.  202007820024).

\bibliographystyle{splncs04}
\bibliography{OM2021}

\begin{thebibliography}{10}
\providecommand{\url}[1]{\texttt{#1}}
\providecommand{\urlprefix}{URL }
\providecommand{\doi}[1]{https://doi.org/#1}

\bibitem{11}
Auer, S., Bizer, C., Kobilarov, G., Lehmann, J., Cyganiak, R., Ives, Z.:
  Dbpedia: A nucleus for a web of open data. In: The semantic web, pp.
  722--735. Springer (2007)

\bibitem{52}
Autayeu, A., Giunchiglia, F., Andrews, P.: Lightweight parsing of classications
  into lightweight ontologies. In: European Conference on Research and Advanced
  Technology for Digital Libraries (ECDL 2010). Glasgow, United Kingdom
  (September 2010)

\bibitem{51}
Bella, G., Giunchiglia, F., McNeill, F.: Language and domain aware lightweight
  ontology matching. Journal of Web Semantics  \textbf{43},  1--17 (2017)

\bibitem{53}
Bella, G., Zamboni, A., Giunchiglia, F.: Domain-based sense disambiguation in
  multilingual structured data. In: DIVERSITY Workshop at ECAI 2016 (2016)

\bibitem{21}
Bulygin, L., Stupnikov, S.A.: Applying of machine learning techniques to
  combine string-based, language-based and structure-based similarity measures
  for ontology matching. In: DAMDID/RCDL. pp. 129--147 (2019)

\bibitem{18}
Cheatham, M., Hitzler, P.: String similarity metrics for ontology alignment.
  In: International semantic web conference. pp. 294--309. Springer (2013)

\bibitem{6}
Chen, G., Zhang, S.: Identifying mappings among knowledge graphs by formal
  concept analysis. In: OM@ ISWC. pp. 25--35 (2019)

\bibitem{32}
Church, K.W.: Word2vec. Natural Language Engineering  \textbf{23}(1),  155--162
  (2017)

\bibitem{42}
Cur{\'e}, O.C., Maurer, H., Shah, N.H., Le~Pendu, P.: A formal concept analysis
  and semantic query expansion cooperation to refine health outcomes of
  interest. BMC medical informatics and decision making  \textbf{15}(1), ~1--6
  (2015)

\bibitem{26}
Euzenat, J., Ferrara, A., Hollink, L., Isaac, A., Joslyn, C., Malais{\'e}, V.,
  Meilicke, C., Nikolov, A., Pane, J., Sabou, M., et~al.: Results of the
  ontology alignment evaluation initiative 2009  (2010)

\bibitem{13}
Euzenat, J., Meilicke, C., Stuckenschmidt, H., Shvaiko, P., Trojahn, C.:
  Ontology alignment evaluation initiative: six years of experience. In:
  Journal on data semantics XV, pp. 158--192. Springer (2011)

\bibitem{29}
Euzenat, J., Shvaiko, P., et~al.: Ontology matching, vol.~18. Springer (2007)

\bibitem{45}
F{\"a}rber, M., Ell, B., Menne, C., Rettinger, A.: A comparative survey of
  dbpedia, freebase, opencyc, wikidata, and yago. Semantic Web Journal
  \textbf{1}(1), ~1--5 (2015)

\bibitem{22}
Faria, D., Pesquita, C., Santos, E., Palmonari, M., Cruz, I.F., Couto, F.M.:
  The agreementmakerlight ontology matching system. In: OTM Confederated
  International Conferences" On the Move to Meaningful Internet Systems". pp.
  527--541. Springer (2013)

\bibitem{15}
Finch, S.R.: Mathematical constants. Cambridge university press (2003)

\bibitem{48}
Fumagalli, M., Daqian, S., Giunchiglia, F.: Ranking schemas by focus:a
  cognitively-inspired approach. In: 26th International Conference on
  Conceptual Structures (2021)

\bibitem{10}
Ganter, B., Wille, R.: Formal concept analysis: mathematical foundations.
  Springer Science \& Business Media (2012)

\bibitem{46}
Giunchiglia, F., Autayeu, A., Pane, J.: S-match: an open source framework for
  matching lightweight ontologies. Semantic Web  \textbf{3}(3),  307--317
  (2012)

\bibitem{9}
Giunchiglia, F., Fumagalli, M.: Entity type recognition--dealing with the
  diversity of knowledge. In: Proceedings of the International Conference on
  Principles of Knowledge Representation and Reasoning. vol.~17, pp. 414--423
  (2020)

\bibitem{12}
Giunchiglia, F., Zaihrayeu, I., Kharkevich, U.: Formalizing the get-specific
  document classification algorithm. In: International Conference on Theory and
  Practice of Digital Libraries. pp. 26--37. Springer (2007)

\bibitem{7}
Jim{\'e}nez-Ruiz, E., Grau, B.C.: Logmap: Logic-based and scalable ontology
  matching. In: International Semantic Web Conference. pp. 273--288. Springer
  (2011)

\bibitem{34}
Kabir, F., Siddique, S., Kotwal, M.R.A., Huda, M.N.: Bangla text document
  categorization using stochastic gradient descent (sgd) classifier. In: 2015
  International Conference on Cognitive Computing and Information Processing
  (CCIP). pp.~1--4. IEEE (2015)

\bibitem{39}
Lenzerini, M.: Data integration: A theoretical perspective. In: Proceedings of
  the twenty-first ACM SIGMOD-SIGACT-SIGART symposium on Principles of database
  systems. pp. 233--246 (2002)

\bibitem{3}
Lonsdale, D., Embley, D.W., Ding, Y., Xu, L., Hepp, M.: Reusing ontologies and
  language components for ontology generation. Data \& Knowledge Engineering
  \textbf{69}(4),  318--330 (2010)

\bibitem{24}
Nezhadi, A.H., Shadgar, B., Osareh, A.: Ontology alignment using machine
  learning techniques. International Journal of Computer Science \& Information
  Technology  \textbf{3}(2), ~139 (2011)

\bibitem{36}
Ng, A.Y., Jordan, M.I.: On discriminative vs. generative classifiers: A
  comparison of logistic regression and naive bayes. In: Advances in neural
  information processing systems. pp. 841--848 (2002)

\bibitem{33}
Pal, M.: Random forest classifier for remote sensing classification.
  International journal of remote sensing  \textbf{26}(1),  217--222 (2005)

\bibitem{31}
Palmer, M., Wu, Z.: Verb semantics and lexical zhibiao w u. In: Proceedings of
  the 32nd Annual Meeting of the Association for Computational Linguistics, Las
  Cruces, New Mexico. pp. 133--138 (1994)

\bibitem{38}
Patro, S., Sahu, K.K.: Normalization: A preprocessing stage. arXiv preprint
  arXiv:1503.06462  (2015)

\bibitem{37}
Pour, N., Algergawy, A., Amini, R., Faria, D., Fundulaki, I., Harrow, I.,
  Hertling, S., Jim{\'e}nez-Ruiz, E., Jonquet, C., Karam, N., et~al.: Results
  of the ontology alignment evaluation initiative 2020. In: Proceedings of the
  15th International Workshop on Ontology Matching (OM 2020). vol.~2788, pp.
  92--138. CEUR-WS (2020)

\bibitem{16}
Rios-Alvarado, A.B., Lopez-Arevalo, I., Sosa-Sosa, V.J.: Learning concept
  hierarchies from textual resources for ontologies construction. Expert
  Systems with Applications  \textbf{40}(15),  5907--5915 (2013)

\bibitem{35}
Safavian, S.R., Landgrebe, D.: A survey of decision tree classifier
  methodology. IEEE transactions on systems, man, and cybernetics
  \textbf{21}(3),  660--674 (1991)

\bibitem{2}
Shvaiko, P., Euzenat, J.: Ontology matching: state of the art and future
  challenges. IEEE Transactions on knowledge and data engineering
  \textbf{25}(1),  158--176 (2011)

\bibitem{8}
Sleeman, J., Finin, T., Joshi, A.: Entity type recognition for heterogeneous
  semantic graphs. AI Magazine  \textbf{36}(1),  75--86 (2015)

\bibitem{44}
Stumme, G., Maedche, A.: Fca-merge: Bottom-up merging of ontologies. In: IJCAI.
  vol.~1, pp. 225--230 (2001)

\bibitem{19}
Sun, Y., Ma, L., Wang, S.: A comparative evaluation of string similarity
  metrics for ontology alignment. Journal of Information \&Computational
  Science  \textbf{12}(3),  957--964 (2015)

\bibitem{30}
Yujian, L., Bo, L.: A normalized levenshtein distance metric. IEEE transactions
  on pattern analysis and machine intelligence  \textbf{29}(6),  1091--1095
  (2007)

\bibitem{25}
Zamazal, O., Sv{\'a}tek, V.: The ten-year ontofarm and its fertilization within
  the onto-sphere. Journal of Web Semantics  \textbf{43},  46--53 (2017)

\end{thebibliography}
\end{document}